\documentstyle[prd,aps,epsfig]{revtex}
\begin{document}
\twocolumn[\hsize\textwidth\columnwidth\hsize\csname
@twocolumnfalse\endcsname
\title{\Large \bf Spin foams as Feynman diagrams}

\author{\bf Michael Reisenberger${}^{a}$, Carlo 
Rovelli${}^{b}$ \vskip.1cm}
\address{${}^{a}$\ Imperial College, London SW7, UK.} 
\address{${}^{b}$\ Centre de Physique
Th\'eorique, CNRS Luminy, F-13288 Marseille, France.}
\address{${}^{b}$\ Physics Department, University of Pittsburgh,
Pittsburgh, Pa 15260, USA.} 
\address{${}^{a}$ {\rm michaelr@phys.psu.edu}, 
${}^{b}$ {\rm  rovelli@cpt.univ-mrs.fr}} 
\date{\today}

\maketitle

\begin{abstract} 
\noindent It has been recently shown that a certain non-topological
spin foam model can be obtained from the Feynman expansion of a field
theory over a group.  The field theory defines a natural ``sum over
triangulations'', which removes the cut off on the number of degrees
of freedom and restores full covariance.  The resulting formulation is
completely background independent: spacetime emerges as a Feynman
diagram, as it did in the old two-dimensional matrix models.  We show
here that {\em any \/} spin foam model can be obtained from a field
theory in this manner.  We give the explicit form of the field theory
action for an arbitrary spin foam model.  In this way, any model can
be naturally extended to a sum over triangulations.  More precisely,
it is extended to a sum over 2-complexes.

\end{abstract} 
\vskip2pc]
\narrowtext

\section{Introduction}

The spin foam formalism \cite{m1,ba1,iw,rr,r1,ba,ro} is a beautiful
convergence of different approaches to general relativistic --or
background independent-- quantum field theory, and in particular to
quantum gravity.  A surprising variety of theories admit a spin foam
formulation.  Among these are: topological field theories
\cite{t,Ooguri,CraneYetter}; modifications of topological quantum
field theories related to quantum general relativity
\cite{BarrettCrane,Iwasaki,Reisenberg97}; loop quantum gravity
\cite{loop2,loop3}, where spin foams appear as histories of spin
networks \cite{rr,r1}; lattice formulations of covariant theories
\cite{Reisenberg97,FW}; and causal spin-networks \cite{Markopoulou}. 
Spin foams seem therefore to represent a rather general tool for
dealing with background independent quantum field theories, or general
relativistic quantum field theories \cite{jmp}, and thus for
describing quantum spacetime \cite{india}.

The aspect of the spin foam formalism which is less understood is the
``sum over triangulations'' (or the ``fine triangulation limit")
needed to restore full general covariance when the model is not
topological -- that is to say, in the physically interesting case.  In
a recent work \cite{dfkr}, this is problem is addressed in the context
of a specific model, the Barrett-Crane (BC) model \cite{BarrettCrane}. 
It is shown in \cite{dfkr} that the BC model can be obtained from the
perturbative expansion of a field theory over a group.  The BC model
partition function over a given triangulation $\Delta$ is precisely
the Feynman amplitude of a certain Feynman graph determined by
$\Delta$.  The full Feynman expansion of the field theory determines
then a natural generalization of the model to a ``sum over
triangulations'', restoring the infinite number of degrees of freedom
and covariance.  In the present paper, we show that the same
reformulation can be obtained for {\em any\/} spin foam model.  Given
an arbitrary spin foam model, we give here an explicit algorithm for
constructing a field theory whose Feynman expansion gives back the
spin foam model.  This procedure defines an extension to a ``sum over
triangulation'' for any spin foam model.

More precisely, we find that the relevant objects are not
triangulations (hence the quotation marks, above), but weaker
structures: 2-complexes.  This fact confirms previous indications that
2-complexes are the correct objects on which to formulate spin foam
models.  A triangulation determines a 2-complex: the 2-skeleton of its
dual cellular complex; but the converse is not true.  Spin foam models
considered so far have been generally defined over triangulations;
however, they actually depend on the 2-complex only and not the full
triangulation.  The same is true for lattice discretizations of
generally covariant theory.  Furthermore, the spin foam expansion of
canonical loop quantum gravity is in terms of 2-complexes, not
triangulations \cite{rr,r1}.  2-complexes, rather than triangulations,
appear thus to be the natural tool in general relativistic quantum
field theory.

The paper is organized as follows.  In Section II, we give the general
definition of the class of models we consider.  Section III contains
our main result: we show that all models in this class can be obtained
from the Feynman expansion of a field theory, and we give the explicit
algorithm for constructing the action of the field theory from the
vertex amplitude of the spin foam model.  In Section IV, we connect
our formalism to lattice gauge theory.  In particular, following
\cite{Reisenberg97}, we show how each spin foam model can be viewed as
a generally covariant version of a lattice gauge theory.  From this
perspective, the sum over 2-complexes determined by the field theory
is a way of implementing the limit in which the cutoff induced by the
triangulation is removed.  In section V we indicate how to extend our
result to more complex models, and we present some conclusive remarks.

\section{Spin foam models}

We consider models defined by the formal sum
\begin{equation}
 Z =  \sum_{J}\ N(J) 
 \sum_{c}\ \prod_{f\in J} {\rm dim}\, a_{f} \ 
    \prod_{v\in J} A_v(c).  
    \label{eq:Z}
\end{equation}
$J$ is a 2-complex.  A two-complex is a (combinatorial) set of
elements called ``vertices'' $v$, ``edges'' $e$ and ``faces'' $f$, and
a boundary relation among these, such that an edge is bounded by two
vertices, and a face is bounded by a cyclic sequence of contiguous
edges (edges sharing a vertex).

Given a triangulation $\Delta$ of a four dimensional manifold, a
2-complex $J(\Delta)$ is defined as the 2-skeleton of the cellular
complex dual to $\Delta$.  That is, we can identify the vertices $v$
with the 4-simplices of $\Delta$, the edges $e$ with the tetrahedra
and the faces $f$ with the triangles of $\Delta$.  Notice that each
vertex of $J(\Delta)$ bounds precisely five edges and ten faces, and
each edge bounds precisely four faces.  We say that vertices of this
form are 5-valent, and that edges of this form are 4-valent.

The sum in (\ref{eq:Z}) is over all combinatorially
inequivalent 2-complexes, whether or not they come from a
triangulation.  For simplicity, we begin by assuming that
the vertices of $J$ are 5-valent and the edges are 4-valent. 
(That is, we assume that $N(J)$ vanishes unless these
conditions are satisfied).  This restriction is not really
necessary for what follows, and we shall later indicate how
to drop it, but it simplifies presentation substantially.

The ``color'' $c=\{a_{f}, b_{e}\}$ is an assignment of a
unitary irreducible representation $a_{f}$ of a Lie group
$G$ to each face $f$ of $J$, and the assignment of an
``intertwiner'' $b_{e}$ to each edge $e$ of $J$. 
Intertwiners are defined below.  A ``spin foam'' is a
colored 2-complex, namely a couple $(J,c)$. 

Finally, $A_{v}(c)$ is a given function of the $a_{f}$'s and
the $b_{e}$'s that color the ten faces and the five edges
adjacent the vertex $v$.  For each fixed 5-valent vertex 
$v$, we denote the colors of the five adjacent vertices as
$b^{v}_{i}$ and the colors of the ten adjacent faces as
$a^{v}_{ij}$, with $i<j$.  Indices $i,j,k$ take value
$1,\ldots,5$ throughout the paper.  Then
\begin{equation}
     A_{v}(c) = A(a^{v}_{ij},b^{v}_{k}). 
    \label{eq:A}
\end{equation}
Often in the literature, equation (\ref{eq:Z}) is written
with a weight associated to the edges as well.  However,
since each edge is bounded by two vertices, this weight can
always be absorbed in $A_{v}$.

A specific model is determined by choosing a group $G$, a
vertex function $A(a_{ij},b_{k})$, and the weight of each
2-complex $N(J)$.  Various generalizations are possible.  We
already mentioned that vertices and edges of different
valence can be considered.  Also, the $a$'s and $b$'s may be
representations and intertwiners of a quantum group.  

As mentioned in opening, a surprising number of approaches to quantum
gravity, following very different paths, have converged to
formulations of this kind.  For instance: In loop quantum gravity, the
spin foams $(J,c)$ emerge \cite{rr,r1} as histories of spin networks
\cite{spinnet}, that is, histories of quantum states of the space
geometry \cite{qg,qg2}.  The vertex amplitude $A_v(c)$ is given by the
matrix elements of the hamiltonian constraint, and a spin foam has a
natural interpretation as a discretized quantized spacetime. 
In covariant lattice approaches, the sum over colors is the
integration over group elements associated to links, expressed in a
(``Fourier'') mode expansion over the group.  In this case, as we
shall see in detail in Section IV, the vertex amplitude $A_v(c)$ is a
discretized version of (the exponent of) the lagrangian density
\cite{Reisenberg97,FW}.  In topological field theories, the vertex
function is a natural object in the representation theory of the group
$G$, satisfying a set of identities that assure triangulation
independence \cite{t,Ooguri,CraneYetter}.  Finally, in the
modifications of topological quantum field theories related to quantum
general relativity \cite{BarrettCrane,Iwasaki,Reisenberg97}, the
topological field theory vertex amplitude is altered in order to
incorporate a quantum version of the constraints that reduce the BF
topological field theory \cite{BF} to general relativity \cite{BFGR}.

\subsection{Intertwiners, the Turaev-Ooguri-Crane-Yetter and
the Barrett-Crane models}

Given a face $f$ colored with the representation $a_{f}$, 
we associate to $f$ the Hilbert space $H_{f}= H_{a_{f}}$, the 
Hilbert space on which $a_{f}$ is defined. Given an edge $e$, 
bounded by the faces $f_{1}\ldots f_{4}$, we associate to 
$e$ the Hilbert space
\begin{equation}
    H_{e}= H_{f_{1}} \otimes \ldots  \otimes H_{f_{4}}. 
    \label{eq:He}
\end{equation}
$H_{e}$ decomposes in orthogonal subspaces that transform
according to different representations of $G$.  Let
$H_{e}^{0}$ be the invariant subspace (the trivial
representation subspace).  Pick, once and for all, a basis
$(b^{(1)},\ldots, b^{(n)})$ in $H_{e}^{0}$.  An intertwiner
is an element of this basis.  (The notation does not
explicitly indicate the fact that an intertwiner $b$ depends
on the colors $a_{f_u}$ of its four adjacent faces, in the
sense that it is an element of a Hilbert space determined by
such colors.)

Consider now a vertex $v$ bounded by the five edges
$e_{1}\ldots e_{5}$, in turn bounded by the ten faces
$f_{ij}$.  We associate to $v$ the Hilbert space
\begin{equation}
    H_{v}= H_{e_{1}} \otimes \ldots  \otimes H_{f_{5}}. 
    \label{eq:Hv}
\end{equation}
Notice that $ H_{v}= K \otimes K$, where
\begin{equation}
    K = H_{f_{12}} \otimes \ldots  \otimes H_{f_{45}}. 
\end{equation}
The scalar product on $K$ determines naturally a trace on
$H_{v}$ 
\begin{equation}
    Tr(v\otimes w) = (v,w) \ \ \ \ \ v,w \in K. 
     \label{eq:Tr}
\end{equation}
There is thus a quantity naturally associated to a vertex of
$v$ a colored 2-complex, which is
\begin{equation}
 A^{TOCY}(a_{ij},b_{i}) = Tr(b_{{1}}\otimes \ldots \otimes b_{{5}}). 
 \label{eq:ABC}
\end{equation}

The simplest spin foam model is the
Turaev-Ooguri-Crane-Yetter (TOCY) model
\cite{t,Ooguri,CraneYetter}, which is defined by the
following choices.  First, the group $G$ is chosen to be the
quantum group $SU(2)_{q}$ or the Lie group $SU(2)$.  Second,
the vertex function is $A^{TOCY}$, given in (\ref{eq:ABC}). 
Finally, $N(J)$ vanishes for all $J$ except the single
2-complex $J(\Delta)$ determined by a triangulation $\Delta$
of a given 4d manifold $M$.  One can then prove that with
these choices $Z$ is independent from the triangulation
chosen, and depends on the manifold only.  The sum over
colorings diverges for $SU(2)$ and is finite for
$SU(2)_{q}$.  The model can be obtained as a quantization of
4d $BF$ theory \cite{Ooguri,BF}.  Triangulation independence
reflects the fact that $BF$ theory is topological, namely
has only a finite number of degrees of freedom.  The 3d
version of this model is the celebrated Turaev-Viro state
sum \cite{TuraevViro}; its Lie group version is the
Ponzano-Regge \cite{PonzanoRegge} formulation of 3d quantum
gravity.  In turn, Ponzano Regge quantum gravity is
essentially the covariant version of 3d loop quantum
gravity, as it was early recognized in
\cite{RovelliPonzano}.

We can chose an orthonormal basis in each representation
space $H_{f}$.  Vectors in $H_{f}$ can then be written in
terms of their components $v^{\alpha}, \ \alpha= 1, \ldots,
{\rm dim}(a_{f})$.  Vectors in $H_{e}$ can then be written
as $v^{\alpha_{1}\ldots\alpha_{4}}$, where the four indices
live in the four representations associated to the four
faces that $e$ bounds.  In particular, an intertwiner $b$
has the form $b^{\alpha_{1}\ldots\alpha_{4}}$, and is
invariant under the action of the group
\begin{equation}
R^{a_{1}\alpha_{1}}{}_{\beta_{1}}(g) \ldots
R^{a_{4}\alpha_{4}}{}_{\beta_{4}}(g)\
b^{\beta_{1}\ldots\beta_{4}} =
b^{\alpha_{1}\ldots\alpha_{4}}\ \ \ \ \forall g\in G.
    \label{eq:invariantb}
\end{equation}
where $R^{a\alpha}{}_{\beta}(g)$ is the representation
matrix in the representation $a$ of $G$.  The normalization
and orthogonality conditions on the intertwiners read
\begin{equation}
b^{\alpha_{1}\ldots\alpha_{4}}\ b'{}^{\alpha_{1}\ldots\alpha_{4}}
= \delta_{bb'}.
    \label{eq:normalization}
\end{equation}
(Repeated indices are summed.)  A vector in $H_{v}$ has
twenty indices: two for each representation $a_{ij}$
coloring the faces adjacent to the vertex $v$.  The
trace (\ref{eq:Tr}) simply contracts the pairs of
indices in the same representation.  In particular
\begin{eqnarray}
\lefteqn{A^{TOCY}(a_{ij},b_{k}) =} \hspace{2em} \nonumber \\  
& & b_{{1}}^{\alpha_{12}\alpha_{13}\alpha_{14}\alpha_{15}}\ \ 
b_{{2}}^{\alpha_{12}\alpha_{23}\alpha_{24}\alpha_{25}}\ \ 
b_{{3}}^{\alpha_{13}\alpha_{23}\alpha_{34}\alpha_{35}} \nonumber \\ 
& & b_{{4}}^{\alpha_{14}\alpha_{24}\alpha_{34}\alpha_{45}}\ \ 
b_{{5}}^{\alpha_{15}\alpha_{25}\alpha_{35}\alpha_{45}}.
\label{eq:aaaa}
\end{eqnarray}
If we represent each tensor $b$ as a vertex with four lines
--one line per index-- and we connect lines of indices
summed over (Penrose tensor notation) the right hand side
of the above equation yield a 4-simplex, as in Figure 1. 
More precisely, it yields a graph, which we call
$\Gamma_{5}$ that has five 4-valent nodes and ten links. 
Each link connects two distinct nodes, and the graph is the
one-skeleton of a 5-simplex. 

\begin{figure}
\centerline{{\psfig{figure=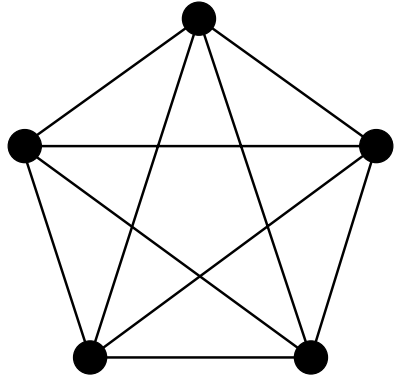,height=5cm}}}
\bigskip \caption{Structure of the vertex function
$A^{TOCY}(a_{ij},b_{k})$: the graph $\Gamma_{5}$.}
\end{figure}

Since in the sequel we have various
expressions with many indices as in (\ref{eq:aaaa}), we
introduce a more compact notation.  We write sequences of
indices with running subindices as indices with subindices
in parenthesis.  That is, for example:
\begin{equation}
    b^{\alpha^{1}_{(i)}} = 
    b^{\alpha^{1}_{2},\alpha^{1}_{3},\alpha^{1}_{4},\alpha^{1}_{5}}.
    \label{eq:notation}
\end{equation}
Then we can write, for instance 
\begin{equation}
    A^{TOCY}(a_{ij},b_{k}) = 
    \prod_{i}b^{\alpha^{i}_{(j)}}\ \ 
    \prod_{i<j}
    \delta_{\alpha^{i}_{j}\alpha^{j}_{i}},
\end{equation}
where, notice, the indices of the two products match. 

Finally, we describe the Barrett-Crane (BC) model
\cite{BarrettCrane}.  The group is $SO(4)$.  The
representations of $SO(4)$ can be labeled by two half
integers $j$ and $j'$ (corresponding to the transformation
properties of the representation under the two $SU(2)$
subgroups).  The representations of $SO(4)$ for which $j=j'$
are called {\em simple}.  Given two simple representations
$a_{1}$ and $a_{2}$, the Hilbert space $H_{12} =
H_{a_{1}}\otimes H_{a_{2}}$ decomposes in a sum over simple
as well as non simple representations.  Let $P_{12}$ be the
projector, defined on $H_{12}$, over its subspace
transforming according to simple representations.  Given an
edge $e$, consider the projector
\begin{equation}
    P_{e} = P_{12}P_{13}P_{14}P_{23}P_{24}P_{34}. 
    \label{eq:P}
\end{equation}
One can prove \cite{prove} that $P_{e}$ projects over a
one-dimensional subspace of $H_{e}$.  Let $b^{BC}$ be the
normalized vector in this one dimensional subspace.  Then
the Barrett-Crane model is defined by the vertex amplitude
\begin{equation}
A^{BC}(a_{ij},b_{k}) = Tr(b_{1}^{BC}\otimes \ldots \otimes
b_{5}^{BC}).  \ \ \prod_{k}\ \delta_{b_{k}b^{BC}_{k}}.
 \label{eq:BC}
\end{equation}
There are a number of indications that this model can be
related to euclidean quantum general relativity.  In fact,
the restriction to simple representations can be viewed as
an implementation of the constraints that reduce $BF$ theory
to general relativity \cite{BFGR}.

Unlikely the TOCY model, the BC model is {\em not\/}
topological.  This is appropriate for a model related to
quantum general relativity because general relativity is
diffeomorphism invariant but has an infinite number of
degrees of freedom \cite{jmp}.  The sum depends therefore
non trivially from the 2-complex (or the triangulation), and
to restore general covariance we have to sum over such
structures.  A natural extension of the model to a sum over
2-complexes was defined in \cite{dfkr}.  The factor $N(J)$
is then given by 
\begin{equation}
    N(J) = \frac{\lambda^{n(J)}}{{\rm Sym}(J)}
    \label{eq:NJ}
\end{equation}
where $n(J)$ and ${\rm Sym}(J)$ are the number of vertices 
and the number of symmetries of $J$ (see \cite{dfkr}).

\section{Field theory}

Given a compact group $G$ and a vertex function
$A(a_{ij},b_{k})$, consider the following function over
$G^{10}$
\begin{equation}
    W(g_{ij}) = \sum_{a_{ij},b_{k}} 
    \overline{\psi}_{a_{ij},b_{k}}(g_{ij}) \ \ A(a_{ij},b_{k}).
    \label{eq:W}
\end{equation}
where $g_{ij}$ is defined for $i<j$ only.  Here
$\psi_{a_{ij},b_{k}}$ is a normalized spin network state on
the graph $\Gamma_{5}$ described above, and in Figure 1,
with nodes colored by the intertwiners $b_{i}$ and links
colored by the representations $a_{ij}$.  The functions
$\psi_{a_{ij},b_{k}}(g_{ij})$ form an a orthonormal basis in
the Hilbert space $L_{2}[G^{10}/G^{5}]$ naturally associated
to this graph.  This is the Hilbert space of a lattice gauge
theory on this graph: the Hilbert space of the Haar square
integrable functions $\psi(g_{ij})$ of ten group elements
$g_{ij}$, invariant under the five gauge transformations
associated to the five vertices of $\Gamma_{5}$
\begin{equation}
    g_{ij} \longrightarrow \rho_{i}g_{ij}(\rho_{j})^{-1}\ \
    \ \ \ (\rho_i\in G). 
    \label{eq:gaugetr}
\end{equation}
Explicitly, the basis is given by 
\begin{equation}
    \psi_{a_{ij},b_{k}}(g_{ij}) 
    = \prod_{i<j} {\rm dim}(a_{ij})\, 
    R^{a_{ij}\alpha_{ij}\alpha_{ji}}\ 
    \prod_{k} b_{k}^{\alpha_{k(i)}}. 
        \label{eq:psi}
\end{equation} 
From $W(g_{ij})$, we define a function of {\rm twenty\/}
group elements $h^i_j$ (with $i\ne j$) by
\begin{equation}
    V(h^i_j) = W\left(h^i_j(h^j_i)^{-1}\right). 
       \label{eq:V}
\end{equation}
(To visualize this step, consider the graph $\Gamma_{5}$. 
Cut its ten links $l_{ij}$ into two parts, which we
denote $l^{i}_{j}$ and $l^{j}_{i}$.  The resulting graph,
$\tilde\Gamma_{5}$ has twenty links, five 4-valent nodes and
ten 2-valent nodes.  Orient each link from the 4-valent node
to the 2-valent node, and associate a group element $h^i_j$
to each link $l^i_j$. See Figure 2.) 

\begin{figure}
\centerline{{\psfig{figure=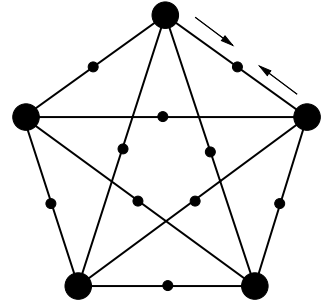,height=4cm}}}
\bigskip \caption{The structure of the vertex function
$V(h^{i}_{j})$: the graph $\tilde\Gamma_{5}$ .}
\end{figure}

Consider then a field theory for a real scalar field
$\phi(h_{1},\ldots, h_{4})$ over $G^{4}$, defined by the
action
 \begin{eqnarray}
  \lefteqn{S[\phi] =} \hspace{1em} \nonumber \\
  & &\int_{G^{4}} dh_{u}\ \phi^{2}(h_{(u)}) 
     + 
     \frac{\lambda}{5!} \int_{G^{20}} dh^{i}_{j}\ 
     V(h^{i}_{j}) \,
     \prod_{k} \phi(h^{k}_{(i)}). 
     \label{eq:S}
 \end{eqnarray}
Indices $u,v$ take value $1,\ldots,4$.  We also assume that
$\phi(h_{1},\ldots, h_{4})$ is $G$ invariant,
\begin{equation} \label{hinv}
	\phi(g_1,g_2,g_3,g_4) = \phi(gg_1,gg_2,gg_3, 
	gg_4), \ \ \ \ \ \ \ \  (\forall g \in G);
\end{equation}
and symmetric 
\begin{equation} \label{perm}
\phi(g_{1},g_{2},g_{3},g_{4})\ =\ 
\phi(g_{\sigma(1)},g_{\sigma(2)},g_{\sigma(3)},g_{\sigma(4)})\ 
\ \ \ \ \ \ \ \  (\forall \sigma),
\end{equation}
where $\sigma$ is a permutation of four elements.  These assumptions
can be dropped by appropriately adjusting the quadratic term in the
action; we keep them for simplicity.  We have then the following
\vskip.1cm
\begin{quote}
{\bf Main result:} \ The formal Feynman perturbation series
of the partition function of this theory
\begin{equation}
Z=\int {\cal D}\phi\, e^{-S[\phi]} 
\label{field}
\end{equation}
is given precisely by
(\ref{eq:Z}), with $N(J)$ given by (\ref{eq:NJ}).
\end{quote}

More in detail, the Feynman graphs of the theory are in 1-1 
correspondence with the two complexes $J$; the momenta of
the field are in 1-1 correspondence with the $a$'s and $b$'s
(they are discrete because the group is compact), and for
each Feynman graph the sum over momenta is precisely the sum
over colorings in (\ref{eq:Z}).

To prove this result, we expand the fields in 
modes over the group, using Peter-Weyl theorem. 
\begin{equation}
\phi(g_{(u)}) = \sum_{a_{u}\alpha_{u}\beta_{u}}\ 
\phi_{a_{(u)}\alpha_{(u)}\beta_{(u)}}
\prod_{u}R^{a_{u}\alpha_{u}\beta_{u}}(g_{u}). 
\end{equation}
It is not difficult to see \cite{dfkr} that gauge invariance
of the field required by equation (\ref{hinv}) implies that 
the field can be written as 
\begin{eqnarray}
\lefteqn{\phi(g_{(u)}) =} \hspace{2em} \nonumber \\
& & \sum_{a_{u}b\alpha_{u}}\! 
\phi_{a_{(u)}b\alpha_{(u)}}\ b^{\beta_{(u)}} \ 
\prod_{u}({\rm
dim}\, a_{u})\ R^{a_{u}\alpha_{u}\beta_{u}}(g_{u}).
\end{eqnarray}
Inserting this expansion in the action, we can read out 
propagator and vertex. The calculation is 
straightforward, and based only on the orthogonality of the 
representation matrices 
\begin{equation}
    \int dg \ \overline{R^{a\alpha}{}_{\beta}(g)}\  
    R^{a'\alpha'}{}_{\beta'}(g) 
    = \frac{1}{{\rm dim}\, a} \ \ \delta_{aa'}\ 
    \delta^{\alpha\alpha'}\ \delta_{\beta\beta'}.
    \label{eq:ortho}
\end{equation}
The propagator turns out to be (recall the field is symmetric) 
\begin{equation}
    P^{a_{u}\alpha_{u};a'_{u}\alpha'_{u}}=
    \sum_{\sigma}
    \prod_{u} \delta^{a_{u}\sigma(a'_{u})}\ \ 
     \ \delta^{\alpha_{u}\sigma(\alpha'_{u})}. 
    \label{eq:propagator}
\end{equation}
And the vertex 
\begin{equation}
    V^{a^i_{j}b_{i}\alpha^i_{j}}=
    A(a^i_j,b_{i})\ 
    \prod_{i<j} \ \delta^{a^i_ja^j_i}\ 
    \delta^{\alpha^i_j\alpha^j_i}. 
\label{vertex}
\end{equation}
The structure of the deltas in the propagator and in the 
vertex is illustrated in Figure 3. 

\begin{figure}
\centerline{$
\hbox{\psfig{figure=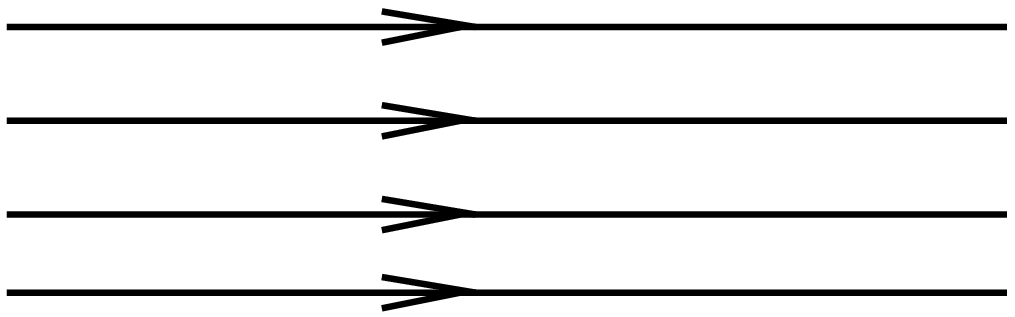,height=1cm}}
\ \ \ \ \ \ \ \ \ 
\hbox{\psfig{figure=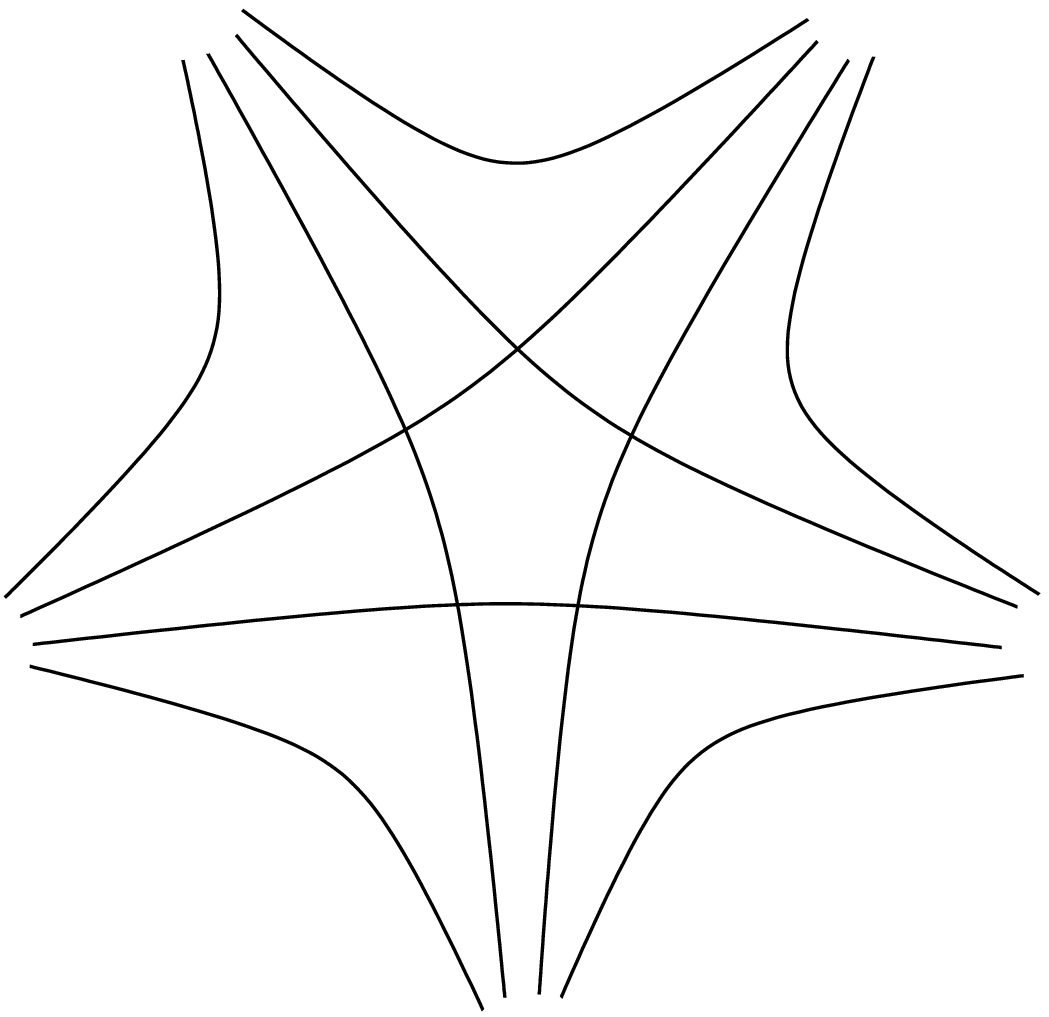,height=4cm}}$}
\bigskip \caption{The structure of the deltas in the propagator and in 
the vertex.}
\end{figure}

A Feynman graph is obtained by taking $n$ vertices and
contracting them with propagators.  Let us call $e_{1}$ one
of these propagators.  Each end of the propagator has four
$\alpha$ indices (see (\ref{eq:propagator})).  The vertex is
five-valent and has twenty $\alpha$ indices (see
(\ref{vertex})).  When contracting one of the $\alpha$
indices of $e_{1}$ with a propagator, this index hits one of
the delta's $\delta^{\alpha^i_j\alpha^j_i}$ in
(\ref{vertex}).  That is, it gets contracted with one of the
$\alpha$ indices of a second propagator.  Call this second
propagator $e_{2}$.  But the propagator $e_{2}$, in turns,
contains a delta function, which connects the index to a
second vertex.  We can thus follow the contraction along the
graph, obtaining a sequence of edges $(e_{1},e_{2},e_{3}
\ldots)$.  Since the graph is finite, the sequence must
close to itself.  In summing the indices of the last delta
we obtain the number $\delta^{\alpha_i^j \alpha_i^j}
\delta_{\alpha_i^j\alpha_i^j}={\rm dim}(a_{ij})$.  (Since,
because of the delta's $a^i_j=a^j_i$, we can forget the
order between $i$ and $j$ and denote this representation as
$a_{ij}, i<j$.)  We can thus drop all the $\alpha$ indices
all together, and add to the sum a factor ${\rm
dim}(a_{ij})$ for each cycle of edges.  The sum over
permutations is then converted in a sum over all ways of
writing cycles over the graph.  But a graph with cycles of
edges is precisely a 2-complex.  The sum is over graph
becomes thus a sum over 2-complexes, in which the
representations $a_{ij}$ label the faces, and the
intertwiners $b_{i}$'s label the edges.  The amplitude is
obtained by multiplying the vertex amplitudes, and the
factor ${\rm dim}\, a_{ij}$ for every face.  Finally, the
weight of each graph is given by standard Feynman-graphology
as the coupling constant to the power of the number of
vertices divided by the symmetry factor of the 2-complex. 
This completes the proof of our main result.

Let us consider an example.  The vertex function for the
$SU(2)$ TOCY model, $V^{TOCY}(g_{ij})$, is obtained by first
inserting $A^{TOCY}(a_{ij},b_{i})$, defined in
(\ref{eq:ABC}) into the equation (\ref{eq:W}).  The result
is that $W^{TOCY}(g_{ij})$ is the distribution with support
on the group elements $g_{ij}=1$ and their gauge
equivalents:
\begin{equation}
    W^{TOCY}(g_{ij}) = \int_{G^{i}} dq_{i} \ 
    \delta\left(q_{i}g_{ij}(q_{j})^{-1}  \right),
    \label{eq:WTOCY}
\end{equation}
where
\begin{equation}
    \int_{G} dg\ \delta(g)\ f(g) = f(1). 
\end{equation}
To prove (\ref{eq:WTOCY}), it is sufficient to integrate its
two sides against all basis elements in the Hilbert space,
and notice that in both cases we get 
\begin{equation}
  \int_{G^{10}} dg_{ij}\ W^{TOCY}(g_{ij}) 
  \psi_{a_{ij},b_{k}}(g_{ij}) 
  =  A(a_{ij},b_{k}). 
    \label{eq:prova}
\end{equation}
Inserting $W^{TOCY}(g_{ij})$ in (\ref{eq:V}) and in the
action (\ref{eq:S}), ten of the twenty integrations can be
performed immediately, because the field is gauge invariant. 
This gives the action
\begin{equation}
    S[\phi] = \int_{G^{4}} dh_{u}\ \phi^{2}(h_{(u)})\ 
    + \ 
    \frac{\lambda}{5!} \int_{G^{10}} dg_{ij}\ 
    \prod_{j} \phi(g_{ij}), 
\end{equation}
where $g_{ji}=(g_{ij})^{-1}$.  This is precisely the Ooguri
action \cite{Ooguri}, or, in three dimensions, the Boulatov
action \cite{Boulatov}, from which the TOCY model was
derived in the first place.

\section{Geometrical interpretation and lattice theory}

We now give a geometrical interpretation to the above
construction.  This interpretation connects the
spin foam formulation with lattice gauge theory.

To this purpose, let us analyze the Feynman expansion
in the ``coordinate'' $g$ space, instead than in the
``momentum'' space of the modes $(a,b)$.  The Feynman
expansion of our field theory is still given as a sum over
Feynman graphs $\Gamma$, with five valent nodes
\begin{equation}
    Z = \sum_{\Gamma} Z(\Gamma), 
\end{equation}
For each graph $\Gamma$, we have four group elements at each
end of each propagator.  Namely, we have four plus four
group elements for each edge $e$ of the graph.  We
denote the group elements at the two ends of the edge $e$ as
$h_{u}^{e}$ and $h'{}_{u}^{e}$, where, we recall $u=1,\ldots
,4$.  We have then immediately
\begin{equation}
    Z(\Gamma) = \int dh^{e}_{u}\ dh'{}_{u}^{e} \  
    \prod_{e}  P(h_{u}^{e},h'{}_{u}^{e})\ 
    \prod_{v} V(h^{i}_{j}),
\end{equation}
where $v$ labels the vertices of the graph, and the twenty
group elements $h^{i}_{j}$ in the argument of $V(\ )$ are
the five times four group elements associated to the four
edges bounded by the vertex $v$.  The propagator
corresponding to the edge $e$ is
\begin{equation}
P(h_{u}^{e},h'{}_{u}^{e}) = 
\sum_{\sigma}\ \prod_{u}\  
\delta(h_{u}^{e},h'{}_{\sigma(u)}^{e}),
    \label{eq:prop}
\end{equation}
so that half of the group integrals can be performed 
immediately, leaving 
\begin{equation}
    Z(\Gamma) = \int dh^e_u \ \ 
    \prod_{v} \ \ \sum_{\sigma_{e}} V(h^e_{\sigma_{e}(u)}). 
\end{equation}
where $\sigma_{e}$ are the permutations of the $u's$ in
$h^{e}_{u}$.  There are now only four group elements
$dh^e_u, u=1,\ldots, 4$ associated to each edge $e$.  

Now, due to equation (\ref{eq:V}), the function $V$ depends
on ten products only, out of the twenty $h^e_u$.  That is,
the twenty group elements in the argument of $V$ get paired. 
Let us number the edges around a vertex as 1 to 5, and, in
each vertex, denote the group elements as $h^i_j,
i,j=1,\ldots, 5$.  Here the upper index $i$ denotes the edge
to which the group element belongs, and the lower index $j$
denotes (four each fixed permutation) the edge to which it
is paired.  Thus $h^i_j$ enters $V(h^i_j)$ only through the
combination $g_{ij}=h^i_j(h^j_i)^{-1}$.  For each given set
of permutations, group elements get paired across the
vertices.  Precisely as we did in momentum space, we can
thus replace the assignment of a fixed set of permutations
with the assignment of all the cycles generated in this
manner.  We identify cycles as faces.  A graph with a full
set of cycles is thus a graph with faces, namely a
2-complex.  Therefore the sum over graphs and the sum over
permutations combine in a sum over 2-complexes $J$, and we
obtain
\begin{equation}
    Z = \sum_{J} Z(J), 
\end{equation}
where the complex amplitude is 
\begin{equation}
    Z(J) = \int dh^e_f \ \ 
    \prod_{v} V(h^e_f).  
\label{mk}
\end{equation}
Where, now the group elements are associated to edges $e$
and adjacent faces $f$.

We can now get to the geometrical interpretation of equation
(\ref{mk}).  Pick a 2-complex $J$.  There is naturally a
lattice $L$ which is, in a sense, ``dual'' to $J$.  To
construct the lattice $L$, imagine that the 2-complex $J$ is
formed by actual surfaces $f$ immersed in a manifold,
joining at the (4-valent) edges $e$ (segments in $M$),
which, in turn, join at the (5-valent) vertices $v$ (points
in $M$).  In other words, let us consider $J$ not as an
abstract combinatorial set, but as 2-dimensional subset of a
manifold.  Now, pick a point $p_{f}$ on each surface $f$ and
a point $p_{e}$ on each edge $e$, and draw an (oriented)
link $l_{f}^{e}$ that goes from each $p_{f}$ to each of the
$p_{e}$ in the edges that bound the face $f$.  The
collection of all these links forms a lattice (a graph),
which we call $L$.  Notice that the nodes of the graph $L$
are of two kinds: the nodes $p_{e}$ are 4-valent, while the
nodes $p_f$ can have arbitrary valence, because a face can
be bound by an arbitrary number of edges.  Each link is
oriented from 4-valent $p_{e}$ node to an n-valent $p_f$
node.

The lattice $L$ has additional structure, deriving from the
vertices of the original 2-complex $J$.  Consider a vertex
$v$ of $J$.  The vertex $v$ is in the boundary of five edges
$e_{i}$ and ten faces $f_{ij}$.  Accordingly, there is a
portion of the lattice $L$ which is ``around'' $v$: the
portion formed by the twenty links $l^{i}_{j} =
l_{f_{ij}}^{e_{i}}$.  We call this portion of the lattice
the ``elementary'' lattice, and denote it as $L_{v}$. 
$L_{v}$ is a small lattice formed by twenty oriented links. 
Each link emerges from one of five 4-valent nodes, and the
links joins in pairs at 2-valent nodes: $l^{i}_{j}$ joins
$l^{j}_{i}$.  (The 2-valent nodes are the n-valent nodes
$p_{f}$ in $L$, of which two links only belong to $L_v$.) 
This is precisely the $\tilde\Gamma_{5}$ graph of Figure 2. 
The full lattice $L$ is formed by putting together many
elementary lattices $L_{v}$.  Two elementary lattices
$L_{v}$ are joined by putting in common, and identifying a
4-valent node $p_{e}$, and its four links $l^{e}_{u}$. 
This, is clearly the operation of joining two vertices with
an edge, seeing in the dual picture.

Let us now consider a lattice gauge theory on the lattice
$L$.  We associate a group element $h^e_f$ to each link
$l^e_f$ of the lattice.  By locality, the action of the
theory $S[h^e_f]$ must be a sum of the discretized
lagrangian density ${{\cal L}_{v}}(h^{i}_{j})$ of each elementary
lattice $L_{v}$.  The partition function is thus
\begin{equation}
    Z(L) = \int dh^e_f\ \ {\rm exp}\{\scriptstyle{\sum}_{v}{\cal 
    L}_v(h^{i}_{j})\}.
\label{mk2}
\end{equation}
which is precisely (\ref{mk}) with 
\begin{equation}
V(h^{i}_{j}) = e^{{{\cal L}_{v}}(h^{i}_{j})} . 
\end{equation} 
Therefore the partition function of our field theory can be
seen as a lattice gauge theory, defined over the lattice
$L$, and then summed over all possible lattices.

The above construction becomes much more clear in the case in which
$J$ is the 2-skeleton of the dual of a 4d triangulation $\Delta$ of a
4d manifold $M$.  In this case the elementary lattices $L_{v}$ are
simply the 4-simplices of $\Delta$.  More precisely, $L_{v}$ is a
graph on the boundary of the 4-simplex.  The boundary of a 4-simplex
is a 3d compact space (a 3-sphere), triangulated by five tetrahedra
$e_{i}$ (dual to the edges), separated by ten triangles $f_{ij}$ (dual
to the surfaces).  The points $p_{e}$ sit in the center of each
tetrahedra.  If we join $p_{e_{1}}$ and $p_{e_{j}}$ with a segment,
the segment must cross a triangle, in a point, which we call
$p_{f_{ij}}$.  Notice that $L_{v}$ is precisely the graph on the
boundary of the 4-simplex on which the ``boundary data'' of reference
\cite{Reisenberg97} are given.  Thus, the potential $V(h^{i}_{j})$ can
be seen as the amplitude for al elementary tetrahedron of the
triangulation, given as a function of the boundary data.  We
illustrate the relation between the elementary lattice
$\tilde\Gamma_{5}$ and the 4-simplex in Figure 4, by going one
dimension down, namely by representing a 3-simplex, that is, a
tetrahedron, and the corresponding elementary lattice
$\tilde\Gamma_{4}$ on its boundary.

\begin{figure}
\centerline{{\psfig{figure=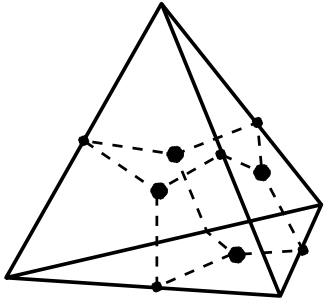,height=4cm}}}
\bigskip \caption{The n-simplex and the elementary lattice
$\tilde\Gamma_{n+1}$ on its boundary, here illustrated in the 
$n=3$ case.}
\end{figure}

We close this section by returning to the example provided by the TOCY
model.  For this model, the potential is the gauge invariant extension
of the delta function on the group.  Thus, the corresponding lattice
gauge theory is obtained by integrating over all group elements such
that the holonomy (the product of the group elements) along each loop
in each $L_{v}$ is the unit in the group.  In the continuum limit,
this is equivalent to integrating over flat connections $A$.  The
restriction to flatness can be obtained with a lagrange multiplier $B$
multiplying the curvature $F$ of $A$, and thus the continuum limit of
the theory is of the form
\begin{equation}
    Z = \int DA\ DB\ \  e^{\imath\int B \wedge F}. 
    \label{eq:BF}
\end{equation}
Which illustrates the relation between the TOCY model and BF
theory \cite{Ooguri}.

\section{Generalizations and conclusions}

It is immediate to generalize our construction to spin foam models
that have vertices of valence different that five.  It suffices to add
a potential term of order $n$ in the fields for each kind of n-valent
vertex of the spin foam model.
It is also immediate to generalize the model to edges of
different valence.  To obtain this, we have to introduce a
different field $\phi(g_{1},\ldots,g_{n})$, with $n$
arguments for each allowed $n$.  In general, the amplitude
of a vertex with $n$ edges and $m$ faces will be determined
by a potential term with $n$ fields depending, altogether,
on $m$ group elements.

In conclusion, we have found that any spin foam model can be obtained
from the Feynman expansion of a suitable field theory on a group
manifold.  In doing that, we have an immediate natural generalization
of the spin foam model to a ``sum over 2-complexes''.  If the model is
topological, then the terms of the sum are independent from the
2-complex, and the sum over 2-complexes is a useless complication. 
However, if the model is not topological, as in models that attempt to
construct quantum theory of the gravitational field, the sum over
2-complexes provides a way to recover the covariance broken by the
choice of a fixed triangulation and to eliminate the artificial cut
off on the number of degrees of freedom introduced by a single
triangulation.

In a quantum gravity model, the sum over colored 2-complexes, or spin
foams, can be seen as a well-defined version of Hawking's sum over
geometries.  Indeed, each colored triangulation can be viewed as a
spacetime with its metric.  Thus, the field theory representation is
the precise 4d analog of the 2d matrix models \cite{2d}, in which a
sum over 2d spacetimes was generated as the Feynman expansion of a
suitable matrix theory.  In fact, the historical path that has lead to
spin foam models from the state sum formulations of topological field
theories started precisely from Boulatov's generalization of the 2d
matrix models to 3 dimensions \cite{Boulatov}.

Canonical quantum general relativity has developed in the loop
representation; as mentioned, remarkably the Feynman's spacetime
representation of loop quantum gravity gives precisely a spin foam
model.  The other way around, the Hilbert space associated to a
canonical formulation of (\ref{eq:Z}) can be represented as the kernel
of a hamiltonian constraint operator (a Wheeler-DeWitt equation) over
a space spanned by a basis of spin networks, precisely as in loop
quantum gravity, and, as in loop quantum gravity, the constraint acts
locally at the nodes of the spin network \cite{ham}.  This remarkable
convergence opens wide possibilities for exploring the theory with the
two complementary tools provided by the covariant and the canonical
theory.  The representation of the sum over spin foams as a field
theory provides a non perturbative handle on the theory, and offers
the intriguing possibility of applying standard quantum field
theoretical machinery.  For instance, conventional renormalization
theory in the field theory context might be useful for dealing with
the potential divergences in the sum.  Quantum field theoretical
methods might also provide helpful in relation to the problem of
extending to the covariant formulation the weave technique
\cite{weave}, namely in identified the coherent quantum states
corresponding to a given classical spactime.

\vskip2cm

We thank Roberto DePietri for discussions and help.  This
work was partially supported by NSF Grant  PHY-9900791.

\end{document}